\documentclass[preprint,12pt]{elsarticle}
\usepackage{xspace}
\newcommand{\Sec}[1]{Section~\ref{#1}}
\newcommand{\EQ}{\begin{equation}}
\newcommand{\EN}{\end{equation}}
\newcommand{\EQA}{\begin{eqnarray}}
\newcommand{\ENA}{\end{eqnarray}}
\newcommand{\eq}[1]{(\ref{#1})}
\newcommand{\Eq}[1]{Eq.~(\ref{#1})}
\newcommand{\Eqs}[2]{Eqs.~(\ref{#1}) and~(\ref{#2})}
\newcommand{\eqs}[2]{(\ref{#1}) and~(\ref{#2})}

\newcommand{\eqss}[2]{(\ref{#1})--(\ref{#2})}
\newcommand{\Section}[1]{Sec.~\ref{#1}}
\newcommand{\App}[1]{Appendix~\ref{#1}}
\newcommand{\Fig}[1]{Figure~\ref{#1}}
\newcommand{\Tab}[1]{Table~\ref{#1}}

\newcommand{\mean}[1]{\overline #1}

\newcommand{\p}{\partial}
\newcommand{\xder}[1]{\frac{\partial #1}{\partial x}}

\newcommand{\timeder}[1]{\frac{\partial #1}{\partial t}}

\def\Rey{\mbox{\rm Re}}

\def\Sh{\mbox{\rm Sh}}
\def\Sr{\mbox{\rm Sr}}

\newcommand{\uu}{{\bf u} {}}
\newcommand{\ttau}{\mbox{\boldmath ${\rm \tau}$} {}}
\newcommand{\SSSS}{\mbox{\boldmath ${\sf S}$} {}}

\usepackage{graphicx}
\journal{Journal of Sound and Vibration}
\begin{document}
\begin{frontmatter}
\title{Non-linear simulations of combustion instabilities with a quasi-1D 
Navier-Stokes code}
\author{Nils Erland L.~Haugen, {\O}yvind Lang{\o}rgen and Sigurd Sannan}
\address{SINTEF Energy Research, NO-7465 Trondheim, Norway}

\begin{abstract}
  As lean premixed combustion systems are more susceptible to combustion 
  instabilities than non-premixed systems, there is an increasing demand 
  for improved numerical design tools that can predict the occurrence of 
  combustion instabilities with high accuracy. The inherent non-linearities in
  combustion instabilities can be of crucial importance, and we here propose 
  an approach in which the one-dimensional Navier-Stokes and scalar transport 
  equations are solved for geometries of variable cross-section. The focus 
  is on attached flames, and for this purpose a new phenomenological 
  model for the unsteady heat release from a flame front is introduced.
  In the attached flame method (AFM) the heat release occurs over the 
  full length of the flame. The non-linear code with the use of the
  AFM approach is validated against results from an experimental study of 
  thermoacoustic instabilities in oxy-fuel flames by Ditaranto and Hals 
  [Combustion and Flame, 146, 493-512 (2006)]. The numerical simulations 
  are in accordance with the experimental measurements and both the 
  frequencies and the amplitudes of the resonant acoustic pressure modes
  are reproduced with good accuracy.
\end{abstract}


\end{frontmatter}

\section{Introduction}

New and stricter emission regulations, particularly on nitrogen oxides, have led to 
the adoption of lean premixed combustion as the primary technology for low-emission 
power generation from gaseous fuels in stationary gas turbines. Homogeneous premixing 
of the fuel and the oxidizer, combined with an ultra-lean operation mode, provide for 
lower combustion temperatures and a drastic reduction in NO$_{\rm x}$ formation,  
accompanied by lower emission of soot and CO. Experience has shown, however, that 
lean premixed combustion systems are more susceptible to combustion instabilities 
than non-premixed systems. The coupling between unsteady heat release 
and acoustic 
pressure oscillations can lead to self-excited oscillations that cause 
unacceptable 
levels of noise and moreover tend to reduce combustion efficiency. These 
thermoacoustic oscillations also have a detrimental effect on the combustor 
equipment that limits the component lifetime and, in the worst case,  can result 
in system failure due to structural damage.

The thermoacoustic instabilities involve a feedback cycle in which the heat 
release rate and acoustic pressure fluctuations are coupled. A stability criterion 
was deduced by Lord Rayleigh \cite{rayleigh}, which states that the pressure 
wave will be amplified and develop an instability if the heat release fluctuations 
are in phase with the acoustic pressure oscillations. Oppositely, the system is stable 
if the heat release rate and acoustic pressure fluctuations are out of phase. In 
general, there are various types of combustion instabilities due to the different 
physical mechanisms that can drive the instability. The underlying driving mechanism 
depends on the combustor geometry, the burner and flame types, and the set-up of the 
fuel and oxidizer feed lines. Thus, an instability due to variation in the equivalence
ratio may arise if the acoustics is able to interact with the upstream flow all the
way to the individual feed lines. Another type of instability occurs if the acoustic
pressure oscillations affect the incoming velocity of the fuel-oxidizer mixture into 
the flame front. Reviews of the acoustically coupled combustion instabilities, as well 
as other instability mechanisms that do not couple directly to the system acoustics, 
are given by Candel \cite{candel_92} and McManus \emph{et al.}~\cite{mcmanus_etal93}.

Combustion instabilities and the interactions between several different physical 
phenomena such as variations in the flow velocities, acoustic pressure fluctuations, 
and heat release are inherently complex by nature. Traditionally, combustion 
system designers have relied on experiments in order to obtain vital information for 
the design of combustors where instabilities can be avoided or controlled. However, 
performing a series of large-scale experiments, in some cases under high pressure to 
create gas turbine operating conditions, is very costly. For this reason the 
development of numerical design tools based on Computational Fluid Dynamics (CFD) 
has become an important endeavor in recent years in order to complement experiments
in the design process. The complexity of the instability mechanisms and their 
interactions, however, places severe demands on both the physical modeling and the 
numerical analysis of current CFD techniques.  Hence, the development of CFD as a 
reliable design and analysis tool with sufficiently high prediction accuracy with 
regard to combustion instabilities is an extremely challenging task.

The most accurate method to study turbulent combusting flows numerically is by 
Direct Numerical Simulation (DNS), where the flow field is solved directly from 
the Navier-Stokes equations~\cite{bell_etal07}. Due to the huge computational cost, 
however, the use of DNS is restricted to simplified problems and turbulent flows 
with relatively low Reynolds number. For high Reynolds number flows in complex 
geometries, as often encountered in industrial applications, DNS is therefore 
prohibitively expensive and is likely to be beyond reach for quite some time. 
A computationally less demanding approach is Large Eddy Simulation (LES), in 
which the large turbulent eddies of the flow are computed explicitly, while the 
smaller eddies are not resolved but modeled using a subgrid scale model. LES is 
well suited to the description of unsteady dynamical phenomena, and there has
been a growing interest in LES in recent years \cite{lartigue_etal04}. But 
even LES is limited due to expensive computational costs, and thus 3D LES 
calculations are non-practical for use on an everyday basis in industrial 
applications.

As a consequence of the complexity of instability mechanisms, much of the 
modeling work on combustion instabilities has focused on simplified models 
to make problems tractable. The development of linear models, in which the 
Navier-Stokes equations and scalar transport equations are linearized, has
been a very useful approach capable of predicting combustion instabilities 
at a qualitative level \cite{mcmanus_etal93}. Thus linear acoustic models 
can predict, at least to some extent, the frequencies of resonant modes and 
their growth rate during a phase of exponential growth. A current practice 
in the modeling of unstable combustion systems is to apply a network model 
where the geometry of the combustor is modeled by a network of acoustic 
elements and a simplified form of the pressure equation is solved. The 
acoustic elements of a network model, also called multiports, correspond 
to various components of the system, \emph{e.g.}, the air or fuel supply, 
the transition between two regions of different cross-section, the outlet 
of the combustor, or the flame itself 
\cite{poinsot_veynante05,keller95,dowling99,paschereit_etal_01,polifke_etal_01}. 
Mathematically, each multiport in the system is represented by a transfer 
function.


Although the predictive scope of linear models is limited, the use of linear 
models for active control of combustion instabilities has been widespread 
\cite{mcmanus_etal93,annaswamy_etal_97}. However, linear models are not able to 
predict the amplitudes of the resonant modes of the statistically stationary state, 
and hence the dominating instabilities of the system cannot be distinguished from the 
less significant ones. Also, the couplings between the instability mechanisms may 
not be accurately represented in linear models and there has therefore been a growing 
interest in active control based on non-linear models \cite{fichera_pagano06}.
When the non-linear Navier-Stokes and scalar transport equations are solved, it 
is possible to predict the shape and the level of the acoustic frequency spectrum 
both during the exponential growth phase and for the statistically stationary state.
This should make for a more accurate and efficient active control of the combustion
instabilities. In a non-linear real space description it is furthermore possible 
to apply more reliable models for the heat release, as discussed in 
\Section{attached_flame_method}. An alternative approach to solving the non-linear real 
space equations is to solve a non-linear wave equation in the frequency 
domain~\cite{culick76a,culick76b}.


In this paper we have chosen an approach where the non-linear Navier-Stokes and 
scalar transport equations are solved for a combusting flow in real space and time, 
as in a DNS or an LES, but in a reduced one-dimensional description. Previous
work on real-space 1D non-linear CFD simulations of combustion instabilities
include the work of Polifke \emph{et al.}~\cite{polifke_etal_01} and Rook 
\cite{rook_01}. Polifke \emph{et al.}~\cite{polifke_etal_01} considered 
thermoacoustic oscillations in a straight duct by introducing a
one-dimensional heat release model for a heat source placed in the duct. From 
time-dependent simulations of the dynamical behavior of the heat source, the
authors were able to obtain the transfer matrix of the heat source which again 
could be investigated in a linear network model. Rook ~\cite{rook_01} applied 
one-dimensional CFD to study the acoustical behavior of a burner-stabilized
flame using a pressure correction method to simulate the flame on a ceramic foam 
burner. We here follow the approach of variable cross-section area introduced by 
Cohen et al.~\cite{cohen_etal_03}, and also used by Prasad and Feng 
\cite{prasad_feng04a,prasad_feng04b}, where the one-dimensional equations are written 
for variable-area geometries in order to simulate three-dimensional geometries. 
The reduction of the governing equations to 1D is valid for relatively simple 
geometries for which the flow can be assumed to be quasi-one-dimensional, and where 
the wavelengths are sufficiently large so that the wave motion is well approximated 
by a plane wave. In this framework we propose a new phenomenological model for the 
unsteady heat release from a flame front. The flame model, here termed the attached 
flame method (AFM), gives a 1D realization of the flame front. Using AFM it 
is demonstrated that the 1D simulations capture both the exponential growth
and the nonlinear statistically stationary state of the acoustic oscillations 
at a computational cost far below that of a corresponding DNS or an LES.

In \Sec{equations} we present the governing equations, and make them 
one-dimensional by introducing a variable cross-section. Boundary conditions 
is discussed in \Sec{boundary}, while the new flame model is described in 
\Sec{flame-model}. Finally, the code is verified in \Sec{results} by 
comparing the simulation results with experimental results from a study of
combustion instabilities in oxy-fuel flames by Ditaranto and Hals 
\cite{ditaranto_hals06}.

\section{The governing equations}
\label{equations}
The governing equations for a turbulent combusting flow are based on the 
conservation of mass, momentum and energy, and the transport equations for 
species mass fractions. The 
conservation of mass is represented by the continuity equation
\EQ
\label{continuity}
\frac{\partial\rho}{\partial t}+\nabla\cdot\rho\,\uu = 0,
\EN
while the conservation of momentum gives the Navier-Stokes equations
\EQ
\label{momentum}
\rho\frac{\partial\uu}{\partial t}+\rho\,\uu \cdot \nabla\uu=
-\nabla p + \nabla\cdot\ttau
\EN
where $\rho$ is the density, $\uu$ is the velocity vector, $p$ is the 
pressure, and $\ttau$ is the viscous stress tensor. For a Newtonian fluid
the stress tensor is given by $\ttau=2\mu\SSSS$, where $\mu$ is the dynamic 
viscosity, and $\SSSS$ is the traceless strain tensor with components
\footnote{The bulk viscosity has been set to zero by Stokes' hypothesis.}
\EQ
S_{ij}=\frac{1}{2}\left(\frac{\p u_i}{\p x_j}+\frac{\p u_j}{\p x_i}
-\frac{2}{3}\delta_{ij}\frac{\p u_k}{\p x_k} \right),
\EN
when we use the Einstein summing convention.
The conservation equation for the energy can be rewritten in terms of the
temperature equation
\EQ
\label{temperature}
\rho\frac{\partial T}{\partial t}+\rho\,\uu\cdot\nabla T=
\frac{1}{c_v}\left(-p\,\nabla\cdot\uu +\nabla\cdot(\lambda\nabla T)+
\dot{q}_v+\dot{q}_c\right),
\EN 
where $T$ is the temperature, $c_v$ is the specific heat at constant volume, 
$\lambda$ is thermal conductivity, $\dot{q}_v=2\mu\SSSS^2$ is the viscous
heating source term, and $\dot{q}_c$ is the heat release rate from
combustion. The equations for the mass fractions $Y_k$ of the 
species $k$ can be written as
\EQ
\label{species_eq}
\rho\timeder{Y_k}+\rho\,\uu\cdot\nabla Y_k=\nabla\cdot(\rho D
\nabla Y_k)+\rho\;\!\omega_k\,;\qquad k=1,\ldots, N_S,
\EN
where $D$ is the mass diffusivity, $\omega_k$ is the chemical reaction rate of 
species $k$, and $N_S$ is the number of species. The above set of equations are 
closed via the ideal gas equation of state
\EQ
\label{ideal_EOS}
p=\rho r T,
\EN
where $r$ is the gas constant of the mixture. The mixture gas constant is
given by $r=R/\overline{m}$, where $R=8.31$~kJ/(kmol K) is the universal gas
constant, and $\overline{m}$ is the mean molar mass. The speed of sound in
the gas mixture is
\EQ
\label{sound_speed}
c_0=\sqrt{\gamma r T},
\EN
where $\gamma=c_p/c_v$ is the specific heat ratio, with $c_p$ the specific
heat constant pressure.

\subsection{The variable cross-section 1D approximation}

In a variable cross-section geometry the combusting flow can be solved by a 
quasi-one-dimensional treatment. The reduced governing equations in one
dimension are presented here. In a quasi-1D description the traceless strain
tensor is reduced to
\EQ
\SSSS={\rm diag}\left(\frac{2}{3},-\frac{1}{3},-\frac{1}{3}\right)
\frac{\p u}{\p x}.
\EN
From this it follows that the viscous force is
\EQA
\label{viscforce}
\nabla\cdot\tau
=\nabla\cdot\left(2\nu\rho\SSSS\right)
=\frac{4}{3}\,\mu\left(\frac{\p^2 u}{\p x^2}+
\frac{\p u}{\p x}\frac{\p \ln\rho}{\p x}
\right),
\ENA
where $\mu$ is the dynamic viscosity. Furthermore, the viscous
heating reduces to
\EQ
\dot{q}_v
=2\mu\SSSS^2
=\frac{4}{3}\mu\left(\frac{\p u}{\p x}\right)^2.
\EN
It should be noted that the factor $4/3$ in the above equations is due to 
the 1D approximation.

The quasi-1D continuity equation for a variable cross-section geometry can be 
written as
\EQ
\label{con}
\timeder{\rho}=-\frac{1}{A}\xder{\rho A u},
\EN
where $A$ is the cross-sectional area. It follows that the one-dimensional
equations for the momentum, the temperature and the species mass fractions are
given by
\EQ
\label{mom}
\frac{\partial u}{\partial t}=-u\frac{\partial u}{\partial x}-
\frac{1}{\rho}\frac{\partial p}{\partial x}+\frac{4}{3}\,\nu 
\left(\frac{\p^2 u}{\p x^2}+
\frac{1}{\rho}\frac{\p u}{\p x}\frac{\p \rho}{\p x}\right)+\frac{F_{f,w}}{\rho},
\EN
\EQ
\label{temp}
\frac{\partial T}{\partial t}=-u\xder{T}+
\frac{1}{\rho c_v}\left(-\frac{p}{A}\xder{uA} +
\xder{}\Big(\lambda\xder{T}\Big)+\dot{q}_v+
\dot{q}_c+\dot{q}_{v,w}\right)
\EN 
and
\EQ
\label{yk}
\timeder{Y_k}=
-u\xder{Y_k}+\frac{1}{\rho}\xder{}\Big(\rho D\frac{\partial Y_k}{\partial x}\Big)
+\omega_k,
\EN
where $F_{f,w}$ represents the viscous force from the walls, and
$\dot{q}_{v,w}$ is the corresponding viscous heating. These terms are added to
the system since the viscous force in \Eq{viscforce} contributes to damping in
the streamwise direction only, \emph{i.e.}, wall effects do not naturally
appear in the one dimensional equations. For more information on the 
viscous force $F_{f,w}$, see \App{wall_drag}. If the viscous wall effects are 
negligible, $F_{f,w}$ and $\dot{q}_{v,w}$ in \Eqs{mom}{temp} can be set to zero.

Numerically, the set of equations \eqss{con}{yk} are solved using an explicit
solver. The spatial discretization is sixth-order finite difference, while third-order 
Runge-Kutta is used for the time stepping.

\section{Boundary conditions}
\label{boundary}
\subsection{Open boundaries}
\label{openboundaries}
We assume that the time varying pressure and velocity field in the system can be 
decomposed as
\EQA
\label{perturbations}
p&=&p_0+p'\nonumber\\
u&=&u_0+u'
\ENA
where subscript 0 denote the mean part while primes denote the fluctuating part.
In addition, the density fluctuations $\rho'$ are given by 
\EQ
\label{perturb}
\rho=\rho_0+\rho',
\EN
where $\rho_0$ is the average density. The acoustic pressure fluctuations are 
described by plane harmonic waves given by
\EQ
\label{pprime}
p'=\hat{p}e^{-i\omega t}=
\left( A^+ e^{ikx}+A^-e^{-ikx}\right)e^{-i\omega t},
\EN
where $k=2\pi/\lambda$ is the wave number, $\omega$ is the angular
frequency, and $A^+$ and $A^-$ denote the amplitudes of the right-moving
and left-moving pressure waves, respectively. Similarly, the velocity fluctuations
are given by
\EQ
\label{uprime}
u'=\hat{u}e^{-i\omega t}=
\frac{1}{\rho_0 c_0}\left( A^+ e^{ikx}-A^-e^{-ikx}\right)e^{-i\omega t}.
\EN

In a long duct with an open exit at $x=0$, the impedance at the exit is 
\EQ
\label{ZR}
Z=\frac{\hat{p}}{\hat{u}}=\rho_0 c_0\left(\frac{A^+ +A^-}{A^+ -A^-}\right)=
\rho_0 c_0\left(\frac{1-R}{1+R}\right),
\EN
where the reflection coefficient $R$ is defined by $R=-A^-/A^+$. Rearranging 
the above equation we get
\EQ
\label{reflection_coeff}
R=\frac{\rho_0 c_0-Z}{\rho_0 c_0+Z}.
\EN
We notice that if $Z=0$ there is perfect reflection and $R=1$. If
$Z=\rho_0 c_0$, corresponding to the characteristic impedance of the
medium, $R=0$ and there is no reflection at the exit. Since 
$p'/u'=\hat{p}/\hat{u}=Z$, we may use the equation of state \eq{ideal_EOS}, 
together with \Eq{ZR}, to write
\EQ
p'=\rho r T-p_0=Z u'=\rho_0 c_0\left(\frac{1-R}{1+R}\right) (u-\mean{u}).
\EN
Solving this equation with respect to the temperature, we obtain
\EQ
\label{temp2}
T=\frac{1}{\rho r}\left(p_0+
\rho_0 c_0 (u-\mean{u}) \left(\frac{1-R}{1+R}\right)\right).
\EN
Thus, at an open boundary the temperature is determined by the expression
\eq{temp2}. 


\subsubsection{Reflection coefficients at open boundaries}
We consider acoustic oscillations in long ducts. If the duct is a circular 
pipe, the characteristic Helmholtz number is defined as $H_n=ka$, where $a$ is 
the radius of the pipe. Thus, the Helmholtz number is small if the pipe radius
is small compared to the acoustic wavelength. It is known that for low
Helmholtz numbers the dominating acoustic modes within a pipe will be
reflected from an open end, \emph{i.e.}, most of the acoustic energy will not 
leave the pipe \cite{levine_schwinger48}. This is normally also the condition 
for combustion instabilities to appear. By assuming that $H_n \ll 1$ and 
applying conservation of mass, it has been shown that the absolute value of
the reflection coefficient at the open end of an unflanged pipe is
\cite{peters_etal_93}
\EQ
\label{Rout}
|R|=1-\frac{1}{2}(ka)^2\,;\qquad ka<0.2.
\EN


For large acoustic velocities $\hat{u}$, non-linear effects at the exit will
no longer be negligible. The dominating non-linearity being the vortex
shedding at the sharp bends of the exit. Peters et al.~\cite{peters_etal_93}
find the acoustic power absorbed by vortex shedding, and non-dimensionalized by 
$P_{\rm norm}=\frac{1}{2}\rho \hat{u}^3\pi a^2$,
to be
\EQ
\label{pvortex1}
P^*_{\rm vortex}
=\frac{P_{\rm vortex}}{P_{\rm norm}}
=\beta \Sr_{ac}^{1/3}
\EN
for the acoustic Strouhal number $\Sr_{ac}=\frac{\omega a}{\hat{u}} \gg 1$, and
\EQ
\label{pvortex2}
P^*_{\rm vortex}
=\frac{2c_d}{3\pi}
\EN
for $\Sr_{ac}\ll 1$.
In the above $c_d=2$ for an unflanged pipe and we have set $\beta=0.5$ in
the simulations. 

If $Re(Z)$ is the real part of the impedance $Z$, the power lost at the 
outlet due to radiation from the pipe exit can be 
found by using \Eq{ZR} and \Eq{Rout} to be
\EQ
\label{Ppi}
P_{\rm rad}=\pi a^2 I=\pi a^2 \frac{1}{2}Re(Z)|\hat{u}|^2,
\EN
where $I$ is the intensity. This gives
\EQ
\label{pradstar}
P^*_{\rm rad}
=\frac{P_{\rm rad}}{\frac{1}{2}\rho \hat{u}^3\pi a^2}
=\frac{1}{4}ka\Sr_{ac}.
\EN
Combining the above equations gives the following expression for the reflection
coefficient as a function of the power loss
\EQ
\label{ref}
R=\frac{\frac{c_0}{\hat{u}}-P_{\rm loss}^*}{\frac{c_0}{\hat{u}}+P_{\rm loss}^*},
\EN
when $P_{\rm loss}^*=P_{\rm vortex}^*+P_{\rm rad.}^*$ is the sum of the
acoustic losses through radiation and vortex shedding.

\subsection{Closed boundaries}

Closed boundaries are understood as acoustically closed boundaries that
reflect acoustic waves. Thus, closed boundaries can either be closed or open
with respect to the mass flow. If there is no inflow (mass flow) across 
the acoustically closed boundary the velocity $u$, together with the derivatives 
$\frac{\partial\rho}{\partial x}$ and $\frac{\partial T}{\partial x}$, are 
set to zero. On the other hand, if there is inflow across the acoustically 
closed boundary, the mass flow is given a constant value
(the velocity $u$ has a non-zero constant value). 

An acoustically closed boundary with inflow can be thought of as wall with 
several small holes, \emph{e.g.}, a porous plate. The holes are so small that 
they do not affect the reflecting abilities of the wall. However, they are 
large enough so that the mass flow entering the domain is significant and 
is allowed to cross the boundary.

By setting $\frac{\partial T}{\partial x}=0$ the closed boundary is adiabatic.
This is not necessarily precisely correct, but the error introduced by this 
assumption is expected to be of minor importance.

\section{The flame model}
\label{flame-model}

In this section we obtain an expression for the combustion heat release term
$\dot{q}_c$ in \Eq{temp}. We begin by describing the well-known $n$-$\tau$ 
formulation and show how it can be integrated into a Navier-Stokes
solver. We then introduce a new phenomenological flame model which we
refer to as the attached flame method (AFM).

\subsection{The $n$-$\tau$ model for unsteady heat release}
As a first approximation we apply the $n$-$\tau$ model, which provides
a global description of the unsteady heat release rate associated with
combustion instabilities. The heat release rate (energy per unit time) 
in the $n$-$\tau$ model is given by \cite{mcmanus_etal93}
\EQ
Q'=\frac{An\rho c_0^{\,2}}{\gamma-1}\,u(x,t-\tau),
\EN
where $A$ is the cross-section of the duct, $n$ is the interaction index
determining the coupling between the velocity and heat release fluctuations,
and $\tau$ is the time lag between these fluctuations. The heat release rate
per volume is
\EQ
\dot{q}_c'=\frac{Q'}{AL_f}=\frac{n\rho c_v\gamma\,T}{L_f}u(x,t-\tau),
\EN
where $L_f$ is the flame length, and we have made use of the relation 
\eq{sound_speed}.

The $n$-$\tau$ model was originally designed for linear wave-equation models 
for which the main focus has been on the heat release fluctuations and their
coupling to the flow-field perturbations. For an implementation of the
$n$-$\tau$ formulation in a CFD code based on the Navier-Stokes equations,
the mean part of the heat release needs to be included as well. Thus, 
the total heat release rate per volume (the mean part plus the fluctuation)
can be written
\EQ
\label{n_tau_eq}
\dot{q}_c=\rho c_v \left(h_c+ \frac{n\gamma\,T}{L_f}\,u(x,t-\tau) \right),
\EN
where $h_c$ is a constant heat source.  Both the constant $h_c$ and the 
unsteady heat release are in this approach non-zero only within the flame.

\subsection{The attached flame method (AFM)}
\label{attached_flame_method}
The $n-\tau$ formulation is primarily designed to be used with linear 
wave-equation models in the frequency domain. Although generalizations to
non-linear applications have been made, the $n-\tau$ model is not optimal 
for use in a non-linear Navier-Stokes solver where the governing equations are
solved in real space and time. Thus, in the $n$-$\tau$ model the heat release
is limited to a point source. In addition, the time lag $\tau$ must be
accounted for. 

More detailed analyses of unsteady heat release from flame fronts have been 
based on studies of the dynamics and shape of anchored premixed flames 
\cite{boyer_quinard90,baillot_etal_92}. Fleifil et al.~\cite{fleifil96} used a 
kinematic model to calculate the transfer function in the linear regime between 
heat release and upstream velocity oscillation of a premixed flame stabilized 
on the rim of a tube. The kinematic approach was later extended by Schuller 
et al.~\cite{schuller_etal_03} by including convective effects of the flow
modulations propagating upstream of the flame. In this paper we present a new 
phenomenological model for the unsteady heat release from an attached flame. 
The model, denoted the attached flame method (AFM), is similar to the analytical 
flame front model \cite{fleifil96,schuller_etal_03} in the sense that a 
real-time differential equation is solved to capture the kinematics of the 
thermoacoustic instabilities. However, while the kinematic model is based on the 
G-equation approach to determine the location of the flame front, in AFM the 
flame position is obtained directly from the equations for the species mass 
fractions. The basic idea is here to project the flame, which is essentially 
two-dimensional, into a one-dimensional description. 

One advantage of the AFM approach, compared to the $n-\tau$ formulation, is that 
for laminar flames no free parameters such as $n$ and $\tau$ need to be determined. 
Another advantage is that the heat release occurs over the full length of the real 
flame, and not just from a single point. In the AFM formulation, the full length 
$L_f$ of the flame is defined as the distance from the inlet, where the flame is 
anchored, to the point where all the fuel is burned. The flame length  $L_f$ is 
therefore a dynamical variable that changes with time as the flow conditions 
change.

The laminar flame speed is known to vary slightly with the position in
a flame. Thus, the laminar flame speed will in general be different at
the base of the flame, in the flame tip, and in the main body of the flame. 
The spatial variations in the flame speed are believed to have only minor effects 
on the flame front, however, and have been neglected in the following.
A known constant laminar flame speed has been assumed in the current formulation,
although this can easily be changed if the effects of a variable flame speed
are expected to be significant. 

\begin{figure}[!ht]\centering
\includegraphics[width=0.60\textwidth]{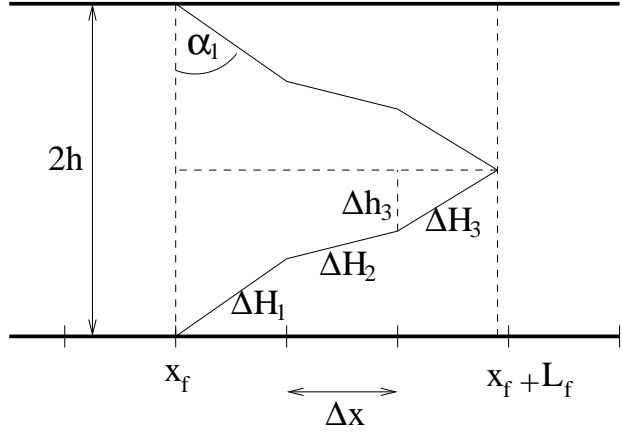}
\caption{
In this duct of height $2h$ there is a flame holder at the position $\rm x_f$ in 
the streamwise direction. The flame front is represented by piecewise straight 
lines with total length $H=2\sum \Delta H_i$, and the length of the flame is 
$\rm L_f$. Every 1D grid cell has length $\rm \Delta x$. 
\label{heating_fig}}
\end{figure}

For a combusting flow in a long duct we assume that the flame front can be 
represented by piecewise straight lines, as shown in \Fig{heating_fig}. 
The total flame surface $\Delta A_i$ within a 1D grid cell is then given
by 
\EQ
\Delta A_i=2 \Delta H_i d,
\EN
where $d$ is the depth of the 1D grid cell, \emph{i.e.}, the size of the grid 
cell in the direction perpendicular to the paper plane. The flame front is here
the interface between the fresh and the burned gas within a given 1D grid cell. 
Since the approach is one-dimensional only, the flame front is not resolved by 
the 1D code but it is evident that the distance from the centerline to the flame 
front, denoted $h_i$, is proportional to the mean mass fraction $Y_{\rm fuel}$ of 
the fuel in the 1D grid cell. This can be written as
\EQ
\label{hi}
\frac{h_i}{h}=\frac{Y_{\rm fuel}}{Y_{\rm fuel,inlet}},
\EN 
where $Y_{\rm fuel,inlet}$ is the mass fraction of the fuel at the inlet. We note
that the limiting cases of $h_i=h$ upstream of the flame anchor at $x_f$ and $h_i=0$ 
downstream of the flame tip are recovered from the above equation. Differentiating 
\Eq{hi} we obtain
\EQ
\frac{d h_i}{d x}=\frac{d Y_{\rm fuel}}{dx}\frac{h}{Y_{\rm fuel,inlet}},
\EN 
which, when setting $\Delta h_i=\Delta x \frac{d h_i}{dx}$, gives
\EQ
\label{eq}
\Delta H_i
=\sqrt{\Delta x^2+\Delta h_i^2}
=\Delta x \sqrt{1+
\left(
\frac{h}{Y_{\rm fuel,inlet}}\frac{dY_{\rm fuel}^i}{dx}
\right)^2}.
\EN
In the above derivation we have used the fact that the mass fraction of the fuel 
outside the flame cone is zero (burned gas), while the mass fraction of the fuel
inside the flame cone equals the mass fraction of the fuel at the inlet (fresh gas).

Since the flame front is always moving into the fresh gas, the reaction rate of the 
fuel within the flame is given by
\EQ
R=\frac{\Delta A_i Y_{\rm fuel,inlet} S_L f_T}{\Delta V},
\EN
where $S_L$ is the laminar flame speed, and $\Delta V$ is the volume of the grid cell.
The reaction rate of the fuel is then given by
\EQ
\label{app2}
R_{\rm fuel}=\left\{
\begin{array}{ll}
0 & x<x_{\rm f}\\
-\frac{2S_L Y_{\rm fuel,inlet}f_T}{h}\sqrt{1+
\left(
\frac{h}{Y_{\rm fuel,inlet}}\frac{dY_{\rm fuel}}{dx}
\right)^2
}& x_{\rm f}\le x \le x_{\rm f}+L_{\rm f}\\
0 & x>x_{\rm f}+L_{\rm f}.
\end{array}\right.
\EN
where $f_T\ge 1$ is a constant accounting for the fact that the real flame speed 
might be larger than the laminar flame speed due to turbulence. In fact, $f_T$, 
quantifying the turbulence in the flame, is the only closure needed in the model. 
In the current work the focus is on laminar problems, however, and the model is here 
validated for laminar or weakly turbulent cases only, for which $f_T=1$.

By construction the AFM does not allow for more than two flame fronts for
any given downstream position. That is; a given flame segment can not
have an angle of more than 90 degrees to the wall. This means that
the model will not be able to describe e.g. vortex roll up, and is therefore
most suited to compact flames. 
This is a natural consequence of the model being one dimensional.

For a chemical reacting system with $N_S$ species, the general equation for
a chemical reaction can be written
\EQ
\sum_{k=1}^{N_S} \nu_k' A_k \rightarrow \sum_{k=1}^{N_S} \nu_k'' A_k,
\EN
where $A_k$ symbolizes the chemical species $k$, and $\nu_k'$ and $\nu_k''$ 
are the stoichiometric coefficients of species $k$ on the reactant and product 
side, respectively. The chemical reaction rate $\omega_k$ for such a system
may be expressed as 
\EQ
\label{omegak}
\omega_k=(\nu_k''-\nu_k')R_{\rm fuel}\frac{m_k}{m_{\rm fuel}},
\EN
where $m_k$ and $m_{\rm fuel}$ is the molar mass of species $k$ and of the
fuel, respectively. 

The energy release within a grid cell is determined by the reaction rate
$R_{\rm fuel}$ and the lower heating value $h_L$ of the unburned mixture.
Thus, the heat release term $\dot{q}_c$ in \Eq{temp} is defined by
\EQ
\label{app3}
\dot{q}_c=R_{\rm fuel} h_L\rho.
\EN

The expressions \eq{omegak} and \eq{app3} are used to close the set of governing 
equations \eqss{con}{yk}.

\subsection{The secondary grid}
\label{secondarygrid}
\begin{figure}[!ht]
\centering
\includegraphics[width=0.70\textwidth]{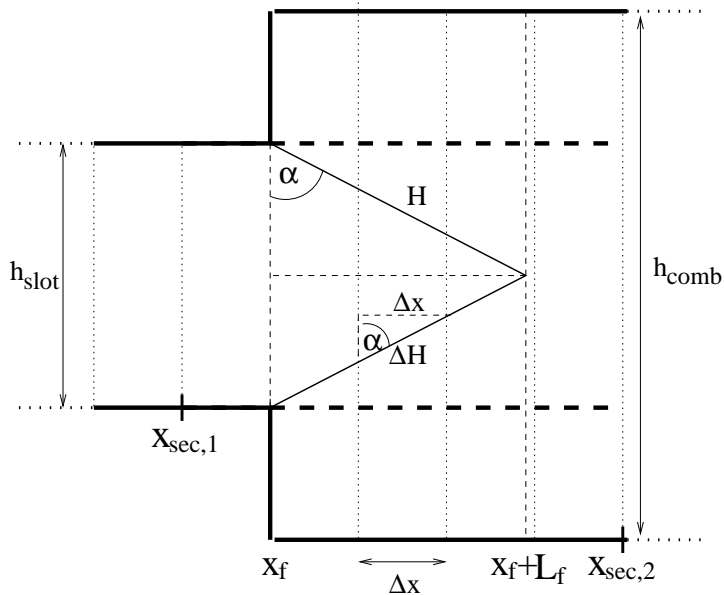}
\caption{The thick solid lines represent the geometry of a slot leading into
a combustion chamber. The thick dashed line corresponds to the upper and
lower boundary of the secondary grid within the chamber. The coordinates 
$\rm x_{sec,1}$ and $\rm x_{sec,2}$ define the boundaries of the secondary
grid in the streamwise direction.
\label{secondary_grid}}
\end{figure}

In many applications of premixed combustion the geometry is such that a narrow 
slot leads into a wider combustion chamber. This is visualized in 
\Fig{secondary_grid} where a slot of height $h_{\rm slot}$ leads into a 
combustion chamber of height $h_{\rm comb}$. In such a geometry it is no 
longer correct to use the same convective velocity for the species convection 
in the jet as in the mean flow. The reason is that the {\it mean} convective 
velocity of the flow within the combustion chamber is lower than the convective 
velocity of the fresh fuel/air jet entering from the slot. Hence, we here solve 
the velocity evolution of the jet using a separate secondary subgrid. The jet 
entering the combustion chamber is visualized by the thick dashed lines in
\Fig{secondary_grid}, which then also represent the upper and lower boundaries
of the secondary grid within the chamber. Note, however, that since the simulation 
tool is formulated in 1D only, no boundary conditions are required at the dashed 
lines. In the streamwise direction the
secondary grid is bounded by the coordinates $\rm x_{sec,1}$ and 
$\rm x_{sec,2}$, as shown in \Fig{secondary_grid}. In this subgrid domain the 
governing equations \eqss{con}{yk} are solved for a constant cross-section $A$
and with no reactions, \emph{i.e.}, $\omega_k$ and $\dot{q}_c$ are both zero. 
This gives the convective velocity $u_{\rm conv}$ to be used in the
convective term of \Eq{mom}. 
The secondary grid is thus applied to only a small fraction 
of the entire computational domain covered by the primary grid, which means
that special care must be taken concerning boundary conditions. At the secondary 
subgrid inlet $\rm x_{sec,1}$ the boundary values are chosen as the 
corresponding instantaneous values of the primary grid at the same location. For 
the subgrid outlet at $\rm x_{sec,2}$ the non-reflecting boundary condition 
given by $R=0$ in \Eq{reflection_coeff} is used.

It should be pointed out that the matching of the secondary and the primary grid 
solutions in the above manner does not have any significant impact on the final 
results as long as both the inlet and the outlet of the secondary grid are outside 
the flame and a non-reflecting boundary condition is used at the secondary grid 
outlet.

\section{Validation of the quasi-1D code}
\label{results}

The non-linear quasi-1D code is here validated against analytical, experimental 
and numerical results. We first compare with analytical results obtained in a 
simplified setup of a flame in a duct. The non-linear code is then validated against 
results from an oxy-fuel study in a lab-scale test rig, both for inert and reactive 
flows.

\subsection{Straight duct}
We here consider a simplified case of a flame in a straight duct for which an 
analytical solution is known when the flame is described by the $n$-$\tau$ model.
For this comparison case the AFM model is not applied since a direct comparison between 
the AFM and the $n$-$\tau$ model is not straightforward, given that various choices 
of $n$ and $\tau$ may lead to very different physical solutions. Thus, the quasi-1D
numerical results are here compared with the analytical solution given by Poinsot and 
Veynante \cite{poinsot_veynante05}, with the $n$-$\tau$ model applied in both 
approaches.

\begin{figure}[!ht]\centering
\includegraphics[width=0.32\textwidth]{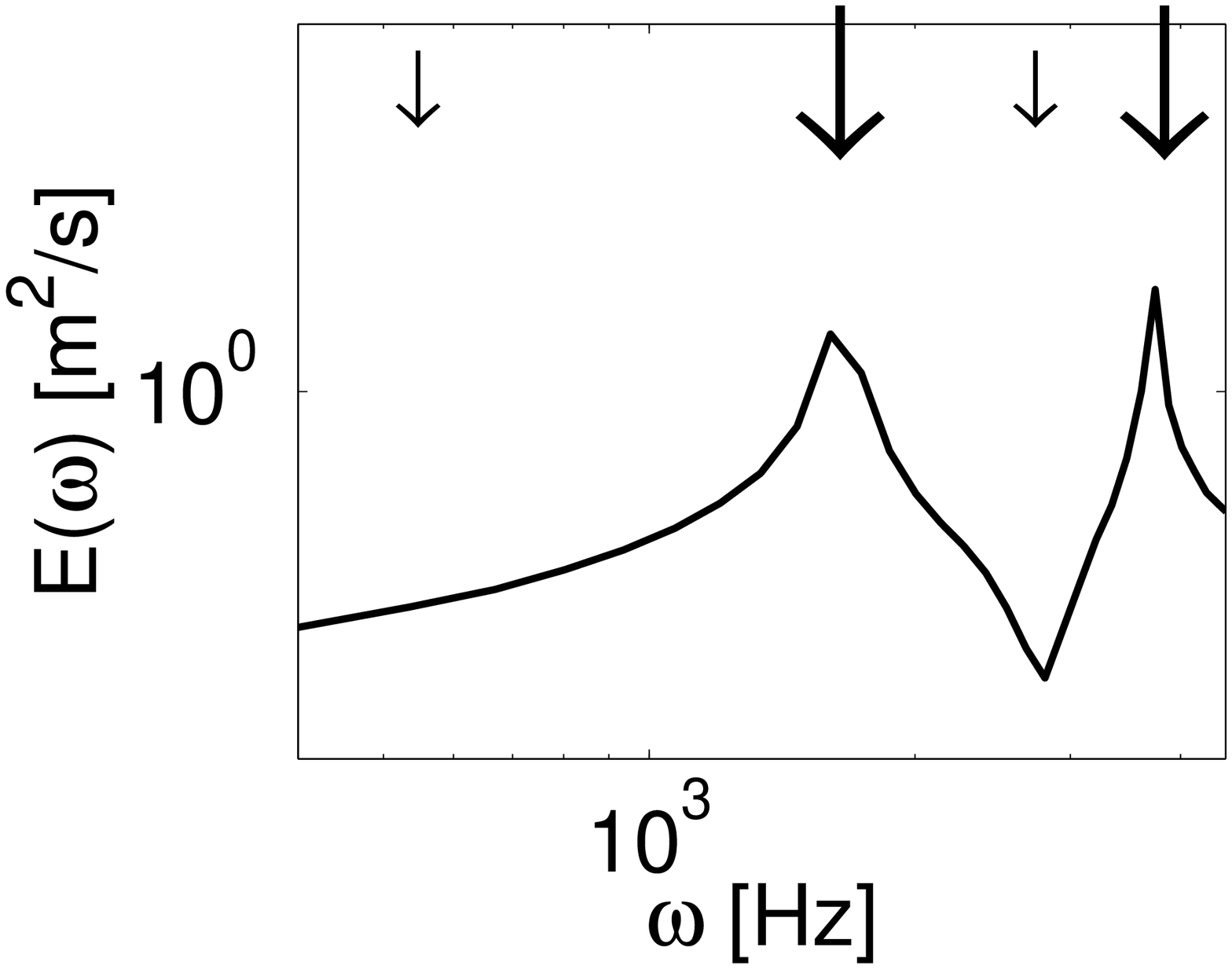}
\includegraphics[width=0.32\textwidth]{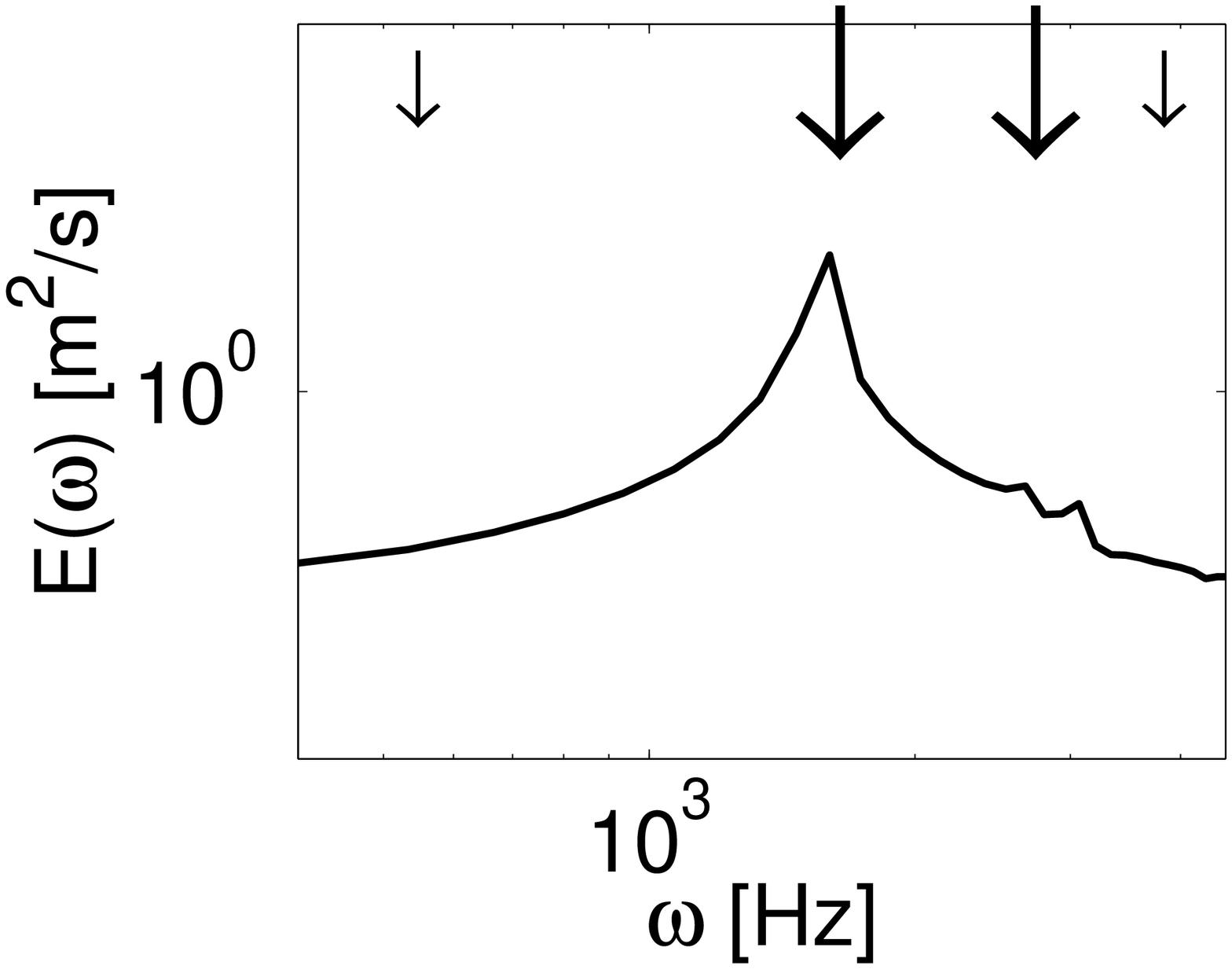}
\includegraphics[width=0.32\textwidth]{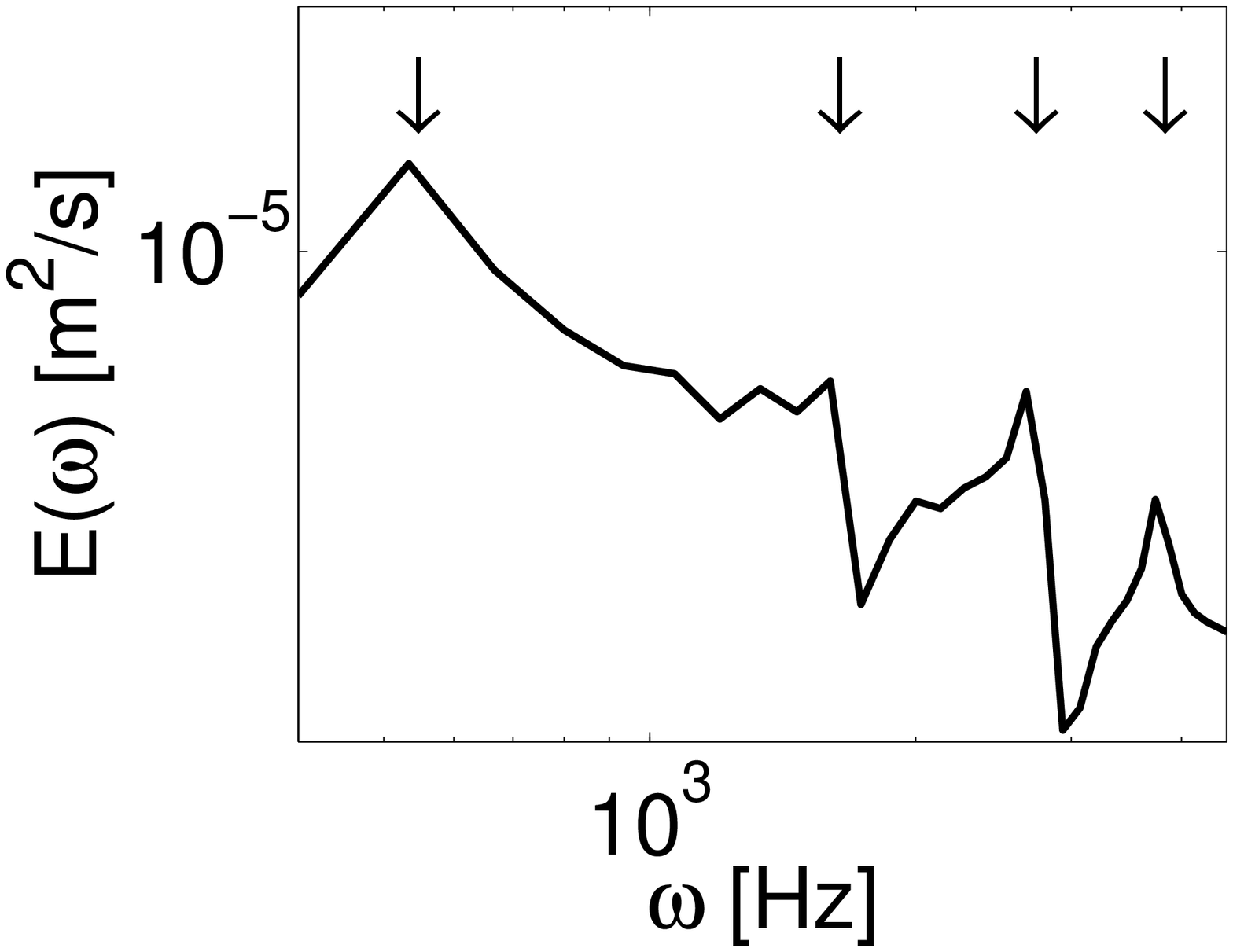}
\caption{Energy spectra for simulations with $\tau=6\times 10^{-4}$~s (left),
$\tau=15\times 10^{-4}$~s (middle)
$\tau=27\times 10^{-4}$~s (right).
\label{spec_tau}}
\end{figure}

The simulated duct is assumed to be a straight pipe of length $2a=1$~m, with the 
flame located in the center of the pipe. For the analytical treatment the heat source 
is a point source, while for the non-linear solver the flame has a length of 8~cm in
the pipe direction. The spatial extension of the flame is required since any gradient 
or object in a spatial code must be resolved by at least a few grid points. The set-up 
is such that air enters at room temperature at the pipe inlet. The constant heat source 
$h_c$ given in \Eq{n_tau_eq} is chosen to be zero such that the mean temperature at the 
outlet equals the inlet temperature. The interaction index $n$ between the velocity and 
the heat release is set to $n=0.25$. We are here primarily interested in the imaginary 
part of the frequency, which for this simplified case is given by
\EQ
Im(\omega_j)=\frac{(-1)^jnc}{8\pi a}\sin(\omega_j\tau),
\EN
where $c$ is the speed of sound, and
\EQ
\omega_j=2\pi f_j=(1+2j)\frac{2\pi c}{8a}.
\EN
In this linear approach the amplitude of a resonant frequency $\omega_j$ is given by 
$A_j=B_j\exp(-i\omega_j t)$, such that the amplitude $A_j$ grows exponentially when 
the imaginary part of $\omega_j$ is positive. If the imaginary part of $\omega_j$ is 
negative, on the other hand, the amplitude of this particular frequency will decay 
exponentially. In \Tab{tab1} the imaginary parts of the first four $\omega_j$'s are 
shown for three different values of $\tau$. For $\tau=6\times 10^{-4}$~s it is evident 
that $\omega_1=1639$~Hz and $\omega_3=3824$~Hz are the unstable frequency modes. In the 
left plot of \Fig{spec_tau} the positions of these two modes are marked by large arrows,
while the smaller arrows correspond to the decaying resonant modes. The solid line
of the plot shows the energy spectrum obtained from the numerical quasi-1D simulation,
and it is seen that the excited modes agree well with the unstable modes as given
by the analytical solution. In the last column of \Tab{tab1} the simulated growth rate 
is also found to be in good agreement with the theoretical values. For 
$\tau=15\times 10^{-4}$~s we note that the unstable frequency modes are given by
$\omega_1=1639$~Hz and $\omega_2=2731$~Hz. In the middle plot of \Fig{spec_tau} we 
observe that the amplitude of $\omega_1$ is much larger than the amplitude of $\omega_2$, 
despite the fact that the imaginary part of $\omega_2$ is larger than that of $\omega_1$. 
This is due to that the initial perturbations are such that $B_1>>B_2$, and that the 
amplitude $A_2$, while still in the linear regime of the simulation, did not have enough 
time to catch up with $A_1$. This is also reflected by the fact that the simulated growth 
rate corresponds to the imaginary part of $\omega_1$. Finally, for $\tau=27\times 10^{-4}$~s 
we find that there are no unstable frequency modes and that all the amplitudes are decaying. 
The resonant modes are nevertheless recovered in the right plot of \Fig{spec_tau}, but 
the amplitudes are very weak (note the scale on the ordinate axis). For this case it should 
be noted that due to quite large uncertainties in the determination of the decay rate, the 
growth rate presented in \Tab{tab1} has relatively large error bars.

\begin{table}[!ht]
  \label{tab1}
  \begin{center}
    \begin{tabular}{l | r r r r | c}
      \hline
      $\tau [10^{-4} s]$&$Im(\omega_0)$&$Im(\omega_1)$&$Im(\omega_2)$&$Im(\omega_3)$
      &Simulated growth rate\\ 
      \hline
      6&  -2.2&   5.8&  -6.9&   5.2&  5.2\\
      15& -5.1&   4.4&   5.7&  -3.6&  4.3\\
      27& -6.9&  -6.6&  -6.2&  -5.4& -5.6\\
      \hline
    \end{tabular}
  \caption{Growth rate of unstable frequencies for different $\tau$'s.
The angular frequencies are $\omega_0$=546~Hz, $\omega_1$=1639~Hz, $\omega_2$=2731~Hz, and
$\omega_3$=3824~Hz.}
  \end{center}
\end{table}

\subsection{Sudden expansion burner}

\begin{figure}[!ht]
\centering\includegraphics[width=1.0\textwidth]{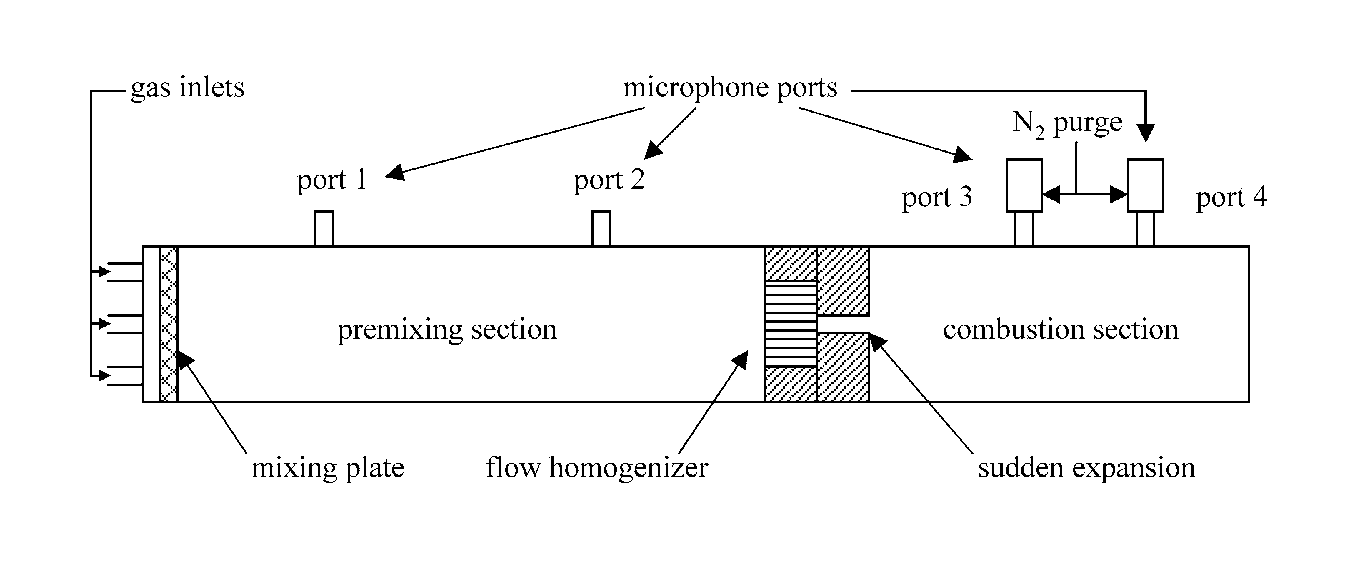}
\caption{Schematic view of the geometry of the oxy-fuel combustion rig used 
in the experimental study of Ditaranto and Hals \cite{ditaranto_hals06}.
\label{geometry}}
\end{figure}

In this section we validate the non-linear quasi-1D code against experimental 
results from the oxy-fuel study of Ditaranto and Hals
\cite{ditaranto_hals06}. A schematic view of the geometry of the lab-scale 
combustion rig is shown in \Fig{geometry}. The set-up consists of a 80 cm long 
premixing section, followed by a 4 cm long flame arrestor and a 10 cm long
slot leading into the 47.9 cm long combustion chamber. The premixing and
combustor sections are square channels with cross-section $s\times s$, where 
$s=5.4$ cm is the inner width of the sections. For the flame arrestor and the 
slot the cross-sections are $s\times d_{arr.}$ and $s \times d_{slot}$, 
respectively, where $d_{arr.}=2$ cm and $d_{slot}=0.5$ cm.

\subsubsection{The cold rig}
\label{coldflow}
In order to validate the non-reactive part of the non-linear code, acoustic
simulations were performed for the geometry of the oxy-fuel rig described  in 
\cite{ditaranto_hals06}, with no flame and only air present in the rig. 
Experimentally, in the cold rig case a loudspeaker was placed at the upstream
end of the premixer, \emph{i.e.}, at the inlet at the left-end side of the 
set-up shown in \Fig{geometry}. The acoustic frequency response of the
loudspeaker was measured with four microphones, and for this procedure there
was no flow in the system. The loudspeaker additionally produced white noise
for which the spectral distribution is not known. The numerical simulations 
can therefore not reproduce the amplitudes of the resonant modes, but merely 
aim at reproducing the resonant frequencies.

\begin{figure}[!ht]
\centering\includegraphics[width=0.70\textwidth]{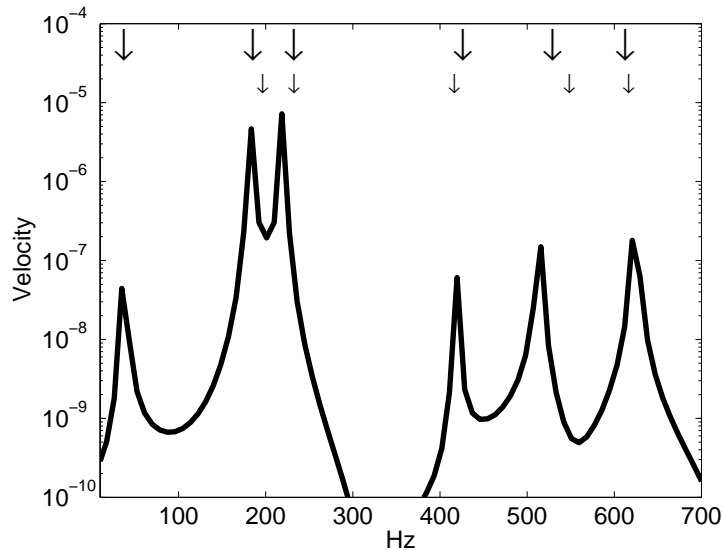}
\caption{The cold flow velocity spectrum from microphone port 3 of the rig
illustrated in \Fig{geometry}. The solid curve shows the spectrum obtained 
from the
non-linear code. The upper arrows indicate the measured resonant frequencies, 
while the lower (smaller) arrows indicate the resonances obtained from the 
linear code. 
\label{velocity_power_spectrum_cold}}
\end{figure}

The velocity spectrum corresponding to the statistically steady state 
solution of the non-linear code is plotted in \Fig{velocity_power_spectrum_cold}. 
Also shown in this figure are the experimentally measured values of the resonant 
frequencies, along with the resonant frequencies obtained with the use 
of the linear code. We note that there is generally a good match between the 
measured resonant modes and those obtained by the numerical simulations.
The resonant peak at $\sim$ 25~Hz corresponds to the low frequency 1/4 mode
of the entire system.

\subsubsection{Oxy-fuel combustion}

\begin{figure}[!ht]
\centering\includegraphics[width=0.70\textwidth]{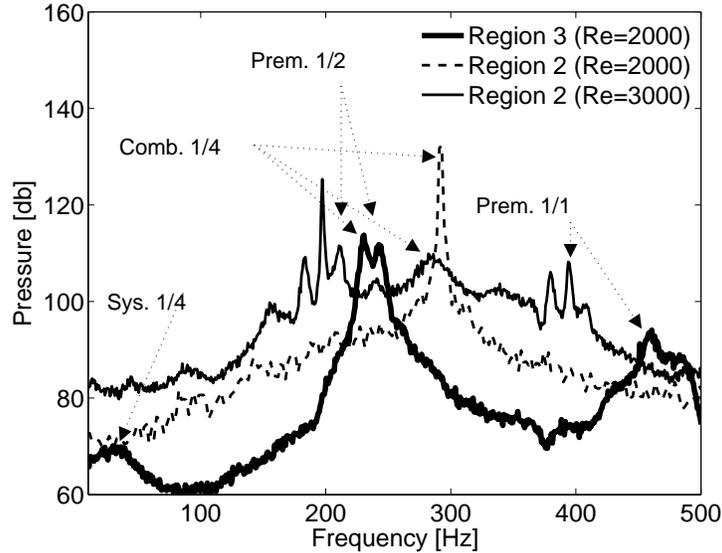}
\caption{The spectral distributions of pressure oscillations in the oxy-fuel
rig as measured by Ditaranto and Hals~\cite{ditaranto_hals06}. The pressure
spectra were recorded at microphone port 3 of the rig in two different
instability regions. In Region 3 (thick solid line) the flame is attached to 
the slot. In Region 2 there are two instability patterns; one (thin solid
line) displaying periodic vortex shedding and another (thin dashed line) being 
an intermediate case with both an attached flame branch and the formation of
periodic vortices. The left arrow in the figure points to the low-frequency
1/4 mode of the entire system, the three next arrows point to the frequency 
peaks of the combustor 1/4 mode, the two upper arrows point to the peaks of 
the premixer 1/2 mode, while the two arrows to the right point to the frequency 
peaks of the 1/1 mode of the premixer. 
\label{compare_exp_spectra}}
\end{figure}

We here validate the full quasi-1D non-linear code using the AFM formulation
against an experimental study of combustion instabilities in oxy-fuel
flames in a sudden expansion test rig. In the oxy-fuel study of Ditaranto and 
Hals~\cite{ditaranto_hals06} four different combustion instability regimes,
referred to as regions, were distinguished. In \Fig{compare_exp_spectra} are 
shown typical spectral distributions obtained in the oxy-fuel experiment for 
two of these regions. The shown pressure spectra were recorded at microphone 
port 3 illustrated in \Fig{geometry}. In what is referred to as Region 3 the 
flame front is attached to the slot of the rig at all times. For Region 2 the
oxy-fuel flame study exhibited two instability patterns; one for which the 
flame is not attached to the slot but follows the formation of periodic large 
vortices, and another corresponding to a hysteresis phenomenon for which there 
is a combination of an attached flame branch and vortex shedding. 

We observe from the spectral distributions of \Fig{compare_exp_spectra} that 
the instability frequencies are quite different for the shown cases. Thus,
for the attached flame of Region 3 the 1/4 mode from the combustion chamber
is at a significantly lower frequency ($\sim 230$ Hz) than the corresponding modes 
\mbox{($\sim 300$ Hz)} for the cases in Region 2. According to Ditaranto and
Hals \cite{ditaranto_hals06}, the instability frequency for the attached flame
case is in fact a combination of the premixer 1/2 mode and the combustion
chamber 1/4 mode. The premixer 1/2 modes for the cases of Region 2 both have
frequencies at \mbox{$\sim 200$ Hz}, although the mode corresponding to the
hysteresis case is not dominant and difficult to observe in 
\Fig{compare_exp_spectra}. Similarly, the premixer 1/1 modes for the cases of 
Region 2 have frequencies at \mbox{$\sim 400$ Hz}, while the corresponding
mode of the attached flame case is at \mbox{$\sim 460$ Hz}. The main
difference between the two instability cases of Region 2 is that the premixer 
1/2 mode is most amplified when the flame follows the shedded vortices
entirely, while the combustor 1/4 mode is dominant when the flame is partially
attached to the slot.


The combustion chamber temperature was not measured by Ditaranto and Hals 
\cite{ditaranto_hals06} and we therefore use the frequency peak of the 
ground mode of the combustor to estimate the average temperature. From 
\Fig{compare_exp_spectra} we note that the 1/4 mode of the combustion chamber 
is at a frequency $f_0\approx 230$~Hz for the attached flame. We first compute
the speed of sound and recall that the combustor has length 47.9 cm. For a 
duct with an open end we also know that the acoustics depends on an end
correction to the duct length. For an open end unflanged pipe, Davies 
et al.~\cite{davies80} have obtained an empirical fit to the end correction 
given by
\EQ
\label{end_corr}
l/a=0.6133-0.1168\,(ka)^2;\qquad ka<0.5,
\EN
where $a$ is the radius of the pipe. For the square combustion duct we set 
$a=s/\sqrt{\pi}=3.0$~cm, corresponding to a pipe of equal cross-section as
the duct. This gives a length correction of $l\approx 1.8$~cm for the ground
mode. Adding this to the combustor length, we find the acoustic 
length of the combustor to be $l_{\rm acoustic}=(0.479+0.018)$ m $=0.497$ m. 
With $f_0=c_0/\lambda$ and $\lambda=4l_{\rm acoustic}$ for the 1/4 wave mode,
this gives the average speed of sound in the combustion chamber
\EQ
\label{sound_vel}
c_0=4f_0l_{\rm acoustic}=458\mbox{ m/s}.
\EN
Using \Eq{sound_speed}, the average temperature in the combustion chamber 
then becomes
\EQ
\label{comb_temp}
T=\frac{c_0^{\,2}}{\gamma r}=720\mbox{ K},
\EN
where $\gamma=1.24$ and $r=235$~J/(kg~K) for the given gas mixture. The
average temperature is thus much lower than the adiabatic flame temperature 
at about 2400~K (when assuming a combustor inlet temperature of 400~K). In 
the following we therefore allow for sufficient cooling in the combustion 
chamber such that the mean temperature downstream of the flame is 720~K.
 
As discussed in \Sec{attached_flame_method} the AFM model is not able to describe
vortex roll up, but is designed to describe attached compact flames.
For the validation of the quasi-1D code a numerical simulation of the case 
for which the flame was attached to the slot has therefore been performed. 
 The case is 
typical of the instability regime of Region 3, and defined by a Reynolds number
$\Rey=2000$, an equivalence ratio $\Phi=0.9$, and a volume fraction of 
42\% O$_2$ in the O$_2$/CO$_2$ oxidant. The corresponding pressure spectrum is 
shown in \Fig{compare_exp_spectra}. In Ditaranto and Hals 
\cite{ditaranto_hals06} the spectral distribution of the pressure oscillations 
is shown in Fig.~7c. From Fig.~1 in \cite{ditaranto_hals06} the laminar flame 
speed is $S_L=0.47$~m/s in the case of 42\% O$_2$ in the oxidant.

In the experimental study the flow rate of fuel and oxidant in the attached 
flame case was $1.3\times 10^{-4}$~kg/s and $1.65\times 10^{-3}$~kg/s,
respectively, corresponding to a mean flow velocity in the premixer of 0.42~m/s.
With $\Rey=2000$ the attached flame is
in the laminar flame regime. Hence, the parameter $f_T$ in \Eq{app2} is set
to $f_T=1$, \emph{i.e.}, the flame speed is equal to the laminar flame speed.

\begin{figure}[!ht]
\begin{center}
\includegraphics[width=6.5cm]{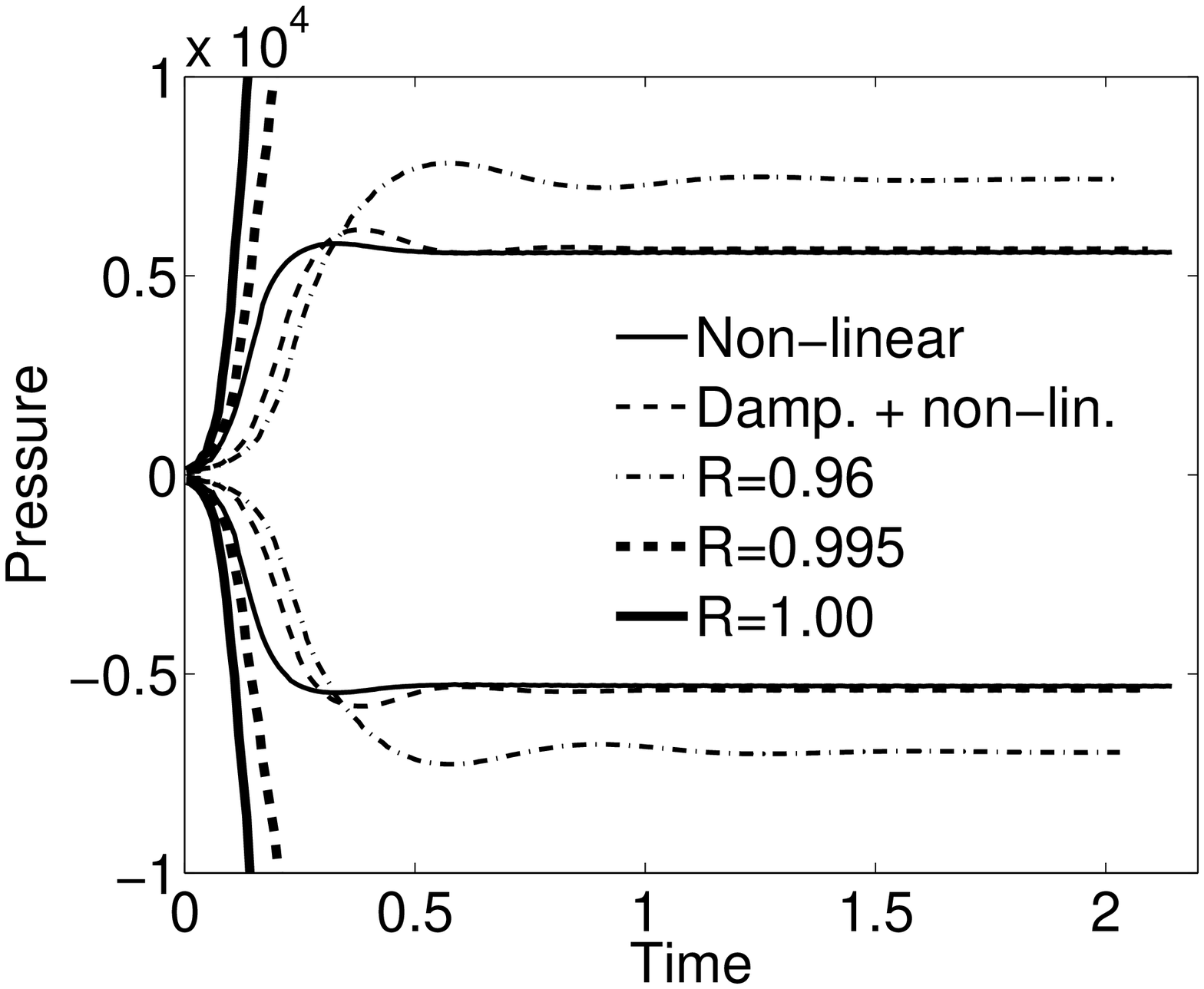}
\includegraphics[width=6.5cm]{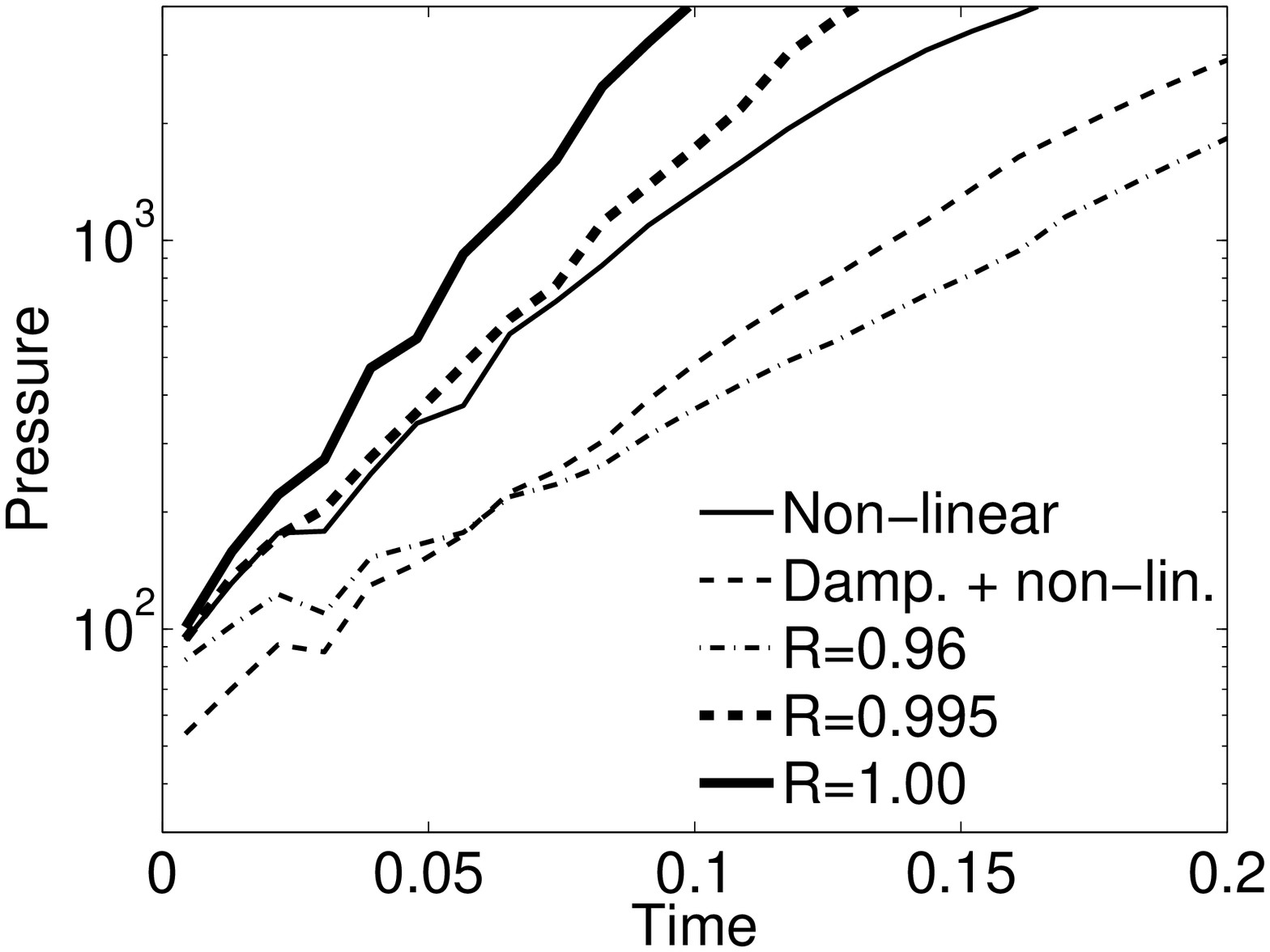}
\caption{Limit cycles for simulations using different reflection properties.
\label{quenching}}  
\end{center}
\end{figure}

The observed thermoacoustic instabilities in the flame experiments are
governed by statistically-stationary limit cycles in which acoustic pressure
variations lead to fluctuations in the flow velocity and heat release. 
The heat release fluctuations, on the other hand, feed back to the
acoustic modes at the same frequency. For the oxy-fuel study the limit cycles
are controlled by a saturation in the heat release caused by acoustic losses 
at the outlet. This can be observed in the left graph of \Fig{quenching},
where the envelope of the pressure amplitudes at microphone port 3 of the rig 
is shown as a function of time for various simulations. We note that when 
there are no acoustic losses, \emph{i.e.}, when the reflection coefficient
$R=1$, the instability grows to infinity. Taking linear acoustic effects into 
account, we obtain from \Eqs{Rout}{sound_vel} a reflection coefficient of 
$R=0.995$ at the dominating frequency \mbox{$f_0=230$ Hz.} The corresponding
instability growth is shown by the thick dashed curve in \Fig{quenching} and 
displays a slightly smaller growth rate than for the $R=1$ case. By including 
both non-linear and radiative losses, the reflection coefficient can be
calculated dynamically from \Eq{ref}. The resulting instability growth is 
shown by the solid curve denoted ``Non-linear'' in \Fig{quenching}. If, in 
addition, viscous damping from the walls of the flame arrestor is taken into 
account, \emph{i.e.}, the term $F_{f,w}$ in \Eq{mom} is non-zero, the growth 
rate is given by the fairly similar dashed curve denoted ``Damp.+non-lin.''. 
(For more details on $F_{f,w}$, see \App{wall_drag}).
Finally, the dashed-dotted curve in \Fig{quenching} shows the pressure
envelope for a simulation for which $R=0.96$.

From the right graph of \Fig{quenching} we observe that the thick dashed curve
and the solid (thin) curve have very similar slopes for small pressure 
variations, while for larger pressure amplitudes the non-linear losses become 
more and more important and the corresponding solid curve levels out at a much 
smaller amplitude than the dashed curve. Comparing the simulation results when 
viscous damping effects were taken into account (dashed curve) with no damping 
effects implemented (solid curve), we observe that the viscous damping only
has little effect on the pressure envelope except for a smaller growth rate 
initially. The simulation using a reflection coefficient of $R=0.96$ was done 
as a comparison case and produced a slightly larger pressure amplitude than 
those indicated for the solid and dashed curves. But the initial growth rate 
produced by the simpler model produced an initial growth rate that was
smaller. From these observations we conclude that it is of crucial importance 
to the numerical simulations that non-linear damping through vortex shedding
at the combustor outlet is included. 
The viscous damping at the walls turns out to be of less
importance for this specific set-up. One issue to be kept in mind, however, is 
that the equations \eqs{pvortex1}{pradstar} for non-linear and radiative
losses, respectively, were deduced for an open end circular pipe. For the 
application of a square duct considered here, additional losses due
\emph{e.g.} to the duct corners might therefore have an impact.

\begin{figure}[!ht]
\begin{center}
\includegraphics[width=8.5cm]{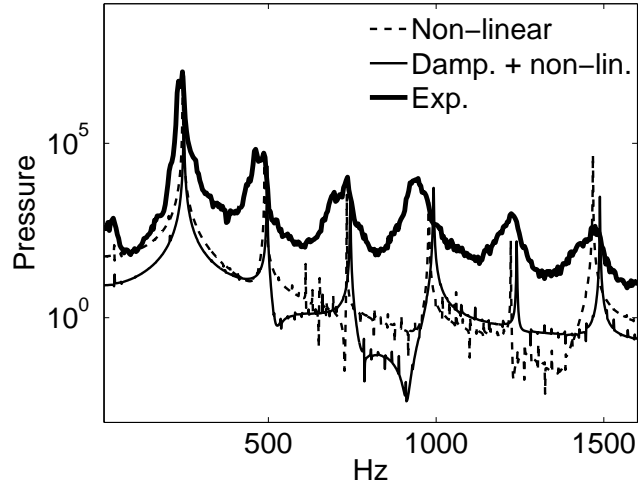}
\caption{Experimental and numerical pressure spectra in the combustion chamber 
at microphone port 3.
\label{pressure_spectrum}}  
\end{center}
\end{figure}

In \Fig{pressure_spectrum} the pressure spectra obtained experimentally
(thick solid curve) and by numerical simulations (thin dashed and solid
curves) are shown. We observe that all the main resonant frequencies of the
experiment are well matched by the simulations. In addition, there is a fairly
good agreement between the experimental and numerical values of the levels of
the resonant peaks of the pressure spectra. However, the levels of the curves 
in between the resonant modes are lower for the numerical simulations than
what was obtained in the experiment. This may be explained by the fact that 
experimental peaks are normally broader than their numerical counterparts, but
the possibility that the difference is due to the  
one-dimensional approximation can clearly also not be excluded. 
It is also interesting to note that the peaks of the two numerical pressure
spectra are well aligned, except for a small difference for the higher 
frequency modes.

\section{Conclusion}

In this paper we first present the quasi-1D Navier-Stokes and scalar transport
equations for a variable cross-section duct. The temperature and species mass 
equations are closed by expressions for the heat release and the reaction
rates obtained from the AFM formulation. The AFM approach has been 
introduced as a new phenomenological flame model in real space and time for 
describing attached flames in a one-dimensional geometry of variable
cross-section. With the use of the secondary subgrid it is shown that
two-dimensional features of a jet entering the combustion chamber is accounted
for. 

The quasi-1D code is non-linear and gives the time evolution of real space 
quantities. This is in contrast to conventional linear wave-equation models
that only give the growth rate of the unstable resonant frequencies, and under
the assumption that a linear representation is adequate. Thus, the quasi-1D 
code is capable of reproducing the non-linear saturation of the acoustic 
oscillations, the so-called limit cycle. In order to achieve the correct 
magnitude of the limit-cycle oscillations, it is crucial to account for the 
acoustic losses at the open end(s) of a duct. This is done by including the 
non-linear effects due to vortex shedding at the sharp bends at the duct exit, 
in addition to the losses due to acoustic radiation from the open end exit.

The non-linear code has been validated by first comparing results from using
the simplified $n$-$\tau$ heat release model with results obtained with the 
$n$-$\tau$ model in a linear wave-equation solver. The code was then validated 
against the oxy-fuel study of Ditaranto and Hals \cite{ditaranto_hals06} in a 
sudden expansion burner. With the application of the AFM heat release 
formulation the numerical simulations were able to reproduce the resonant 
frequencies of the the acoustic pressure spectrum of an oscillating attached 
flame with high accuracy. In addition, the levels of the resonant peaks were 
reproduced quite well. From these findings we conclude that the quasi-1D 
non-linear Navier-Stokes solver with the AFM formulation is a promising tool 
for further studies of combustion instabilities in a variety of cases, 
including jets in variable cross-section ducts or in co-flows.

\section*{Acknowledgments}
This publication forms a part of the BIGCO2 project, 
performed under the strategic Norwegian research program Climit.
The authors acknowledge the partners: StatoilHydro, GE Global Research, 
Statkraft, Aker Kv{\ae}rner, Shell, TOTAL, ConocoPhillips, ALSTOM, 
the Research Council of Norway (178004/I30 and 176059/I30) and 
Gassnova (182070) for their support.

\appendix

\section{Viscous damping}
\label{wall_drag}
The acoustic Strouhal number is defined by $\Sr_{ac}=\omega a/\hat{u}$,
where $\omega$ is the angular frequency of the acoustic oscillations, $a$ is
the radius of the pipe, and $\hat u$ is the amplitude of the acoustic velocity
oscillations. If the acoustic Strouhal number is very small, the boundary
layer develops in a time much smaller than the acoustic period. In that case 
it is a good approximation to assume that the boundary layer is always
developed. Following White \cite{white94}, the viscous force from the walls 
$F_{f,w}$ then is
\EQ
F_{f,w}=-\tau_{w}S_P \Delta x,
\EN
where $\tau_w$ is the wall shear stress, $S_P$ is the duct perimeter, and
$\Delta x$ is the length of a grid cell. The wall shear stress can be
expressed as \cite{white94}
\EQ
\label{tau}
\tau_w=\frac{1}{8}f_D \rho u^2,
\EN
where $f_D$ is the Darcy friction factor. The Darcy friction factor can be
approximated with the expression
\EQ
f_D=\frac{64 f_{\rm prof,1}}{\Rey_{D_h}}
\EN
for laminar flows. For turbulent flows $f_D$ can be obtained from the relation
\EQ
\frac{1}{\sqrt{f_D}}=2.0\log
\left(f_{\rm prof,2}\Rey_{D_h}\sqrt{f_D}\right)-0.8.
\EN
The Reynolds number is based on the hydraulic diameter $D_h=4A/S_P$, where $A$ 
is the cross-section of the duct, such that $\Rey_{D_h}=uD_h/\nu$. In the
above equations, $f_{\rm prof,1}$ and $f_{\rm prof,2}$ are factors whose
values are determined by the shape of the duct. Thus, for a flow in a circular
pipe $f_{\rm prof,1}=f_{\rm prof,2}=1$. For a flow between two parallel plates 
we have $f_{\rm prof,1}=3/2$ and $f_{\rm prof,2}=0.64$. 

It should be noted that the expression \eq{tau} is valid only for fully 
developed flows, see for instance the work by Allam and {\AA}bom
\cite{allam_abom06} for further details. This limitation is most prominent for 
pipes with large cross-sections, \emph{i.e.}, when the Strouhal number
$\Sr_{ac}$ is not much smaller than 1. In a flame trap where the viscous
damping in general is the largest, however, the given expressions should be 
relatively good due to the very small cross-sections of the holes. For a more 
applicable expression for acoustic losses in pipe flows, see the paper by
Disselhorst and Wijngaarden~\cite{disselhorst_wijngaarden_80}.

Following Peters et al.~\cite{peters_etal_93}, for small Helmholtz numbers $ka$ 
and large shear numbers the damping coefficient due to the viscous boundary
layer is given by
\EQ
\label{damping}
\alpha=\frac{\omega}{c_0}\left[\frac{1}{\sqrt{2} \Sh} 
\left(1+\frac{\gamma-1}{\sqrt{\Pr}}\right)+
\frac{1}{\Sh^2}\left(1+\frac{\gamma-1}{\sqrt{\Pr}}-
\frac{\gamma(\gamma-1)}{2\Pr}\right)\right],
\EN
where $\Sh=a\sqrt{\omega/\nu}$ is the shear number and $\Pr$ is the Prandtl
number. The above expression for $\alpha$ has been established under the
assumption of no mean flow in the system. This gives a viscous damping defined 
as 
\EQ
\Delta P'\sim \exp (-\alpha x),
\EN
where $\Delta P'$ is the decrease in the acoustic pressure fluctuations, and
$x$ is the distance traveled by the acoustic waves.

\end{document}